\documentclass[fleqn,10pt]{wlscirep}

\usepackage{amsmath}
\usepackage{amssymb}
\usepackage{graphicx}
\usepackage{textcomp}
\usepackage{multicol}
\usepackage{textcomp}
\usepackage{color}
\usepackage{ulem}

\title{Locating Temporal Functional Dynamics of Visual Short-Term Memory Binding using Graph Modular Dirichlet Energy}

\author[1,2,*]{Keith Smith}
\author[3]{Benjamin Ricaud}
\author[3]{Nauman Shahid}
\author[4]{Stephen Rhodes}
\author[2]{John M. Starr}
\author[5,6,7,8,9]{Augustin Ib\'a\~nez}
\author[2,4,7,10,11]{Mario A. Parra}
\author[1]{Javier Escudero}
\author[3]{Pierre Vandergheynst}
\affil[1]{Institute for Digital Communications, University of Edinburgh, West Mains Rd, Edinburgh, EH9 3FB, UK}
\affil[2]{Alzheimer Scotland Dementia Research Centre, University of Edinburgh, 7 George Square, Edinburgh, EH8 9JZ, UK}
\affil[3]{Signal Processing Laboratory 2, \' Ecole Polytechnique F\' ed\' erale de Lausanne, 1015 Lausanne, Switzerland}
\affil[4]{Human Cognitive Neuroscience and Centre for Cognitive Ageing and Cognitive Epidemiology, Department of Psychology, University of Edinburgh, EH8 9JZ, UK}
\affil[5]{Institute of Translational and Cognitive Neuroscience (INCyT), INECO Foundation, Favaloro University, Buenos Aires, Argentina}
\affil[6]{National Scientific and Technical Research Council (CONICET), Buenos Aires, Argentina}
\affil[7]{Universidad Aut\'onoma del Caribe, Barranquilla, Colombia}
\affil[8]{Department of Psychology, Universidad Adolfo Ib\'a\~nez, Santiago, Chile}
\affil[9]{ARC Centre of Excellence in Cognition and its Disorders, Sydney, Australia}
\affil[10]{Psychology Department, Heriot-Watt University, Edinburgh, EH14 4AS, UK}
\affil[11]{UDP-INECO Foundation Core on Neuroscience (UIFCoN), Diego Portales University, Santiago, Chile}
\affil[*]{k.smith@ed.ac.uk}


\begin{abstract}
  Visual short-term memory binding tasks are a promising early marker for Alzheimer's disease (AD). To uncover functional deficits of AD in these tasks it is meaningful to first study unimpaired brain function. Electroencephalogram recordings were obtained from encoding and maintenance periods of tasks performed by healthy young volunteers. We probe the task's transient physiological underpinnings by contrasting shape only (Shape) and shape-colour binding (Bind) conditions, displayed in the left and right sides of the screen, separately. Particularly, we introduce and implement a novel technique named Modular Dirichlet Energy (MDE) which allows robust and flexible analysis of the functional network with unprecedented temporal precision.  We find that connectivity in the Bind condition is less integrated with the global network than in the Shape condition in occipital and frontal modules during the encoding period of the right screen condition. Using MDE we are able to discern driving effects in the occipital module between 100-140ms, coinciding with the P100 visually evoked potential, followed by a driving effect in the frontal module between 140-180ms, suggesting that the differences found constitute an information processing difference between these modules. This provides temporally precise information over a heterogeneous population in promising tasks for the detection of AD.
\end{abstract}
\begin{document}

\flushbottom
\maketitle

\thispagestyle{empty}

\section*{Introduction}
\begin{figure}[t]
	\centering
	\includegraphics[scale=0.3]{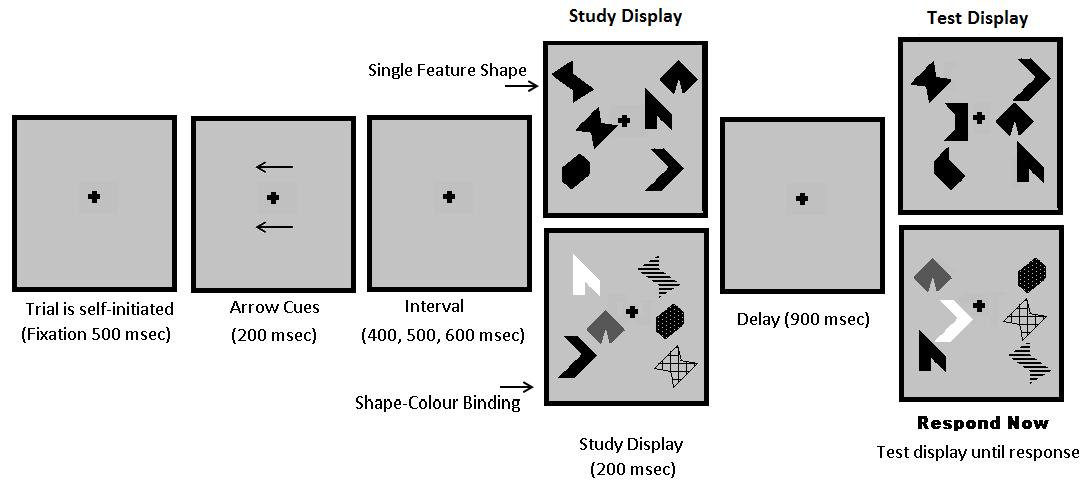}
	\caption{Chronology and design of the Visual Short Term Memory tasks. Arrow cues indicate to the participant the hemisfield being tested before stimulus.}
	\label{Taskfig}
\end{figure}

Visual Short-Term Memory (VSTM) tasks prove promising in the early detection of Alzheimer's Disease (AD). A useful paradigm lies in tasks which are specifically designed to test the ability of participants to store information of either shapes alone (Shape) or shapes with associated colours (Bind) for short-term memory recognition. In this instance it is found that patients are particularly impaired in the Bind condition  \cite{ParAlz1}$^{,}$\cite{ParAlz2}$^{,}$\cite{ParAlz3}. This VSTM task has been found to be both sensitive and specific to early AD \cite{ParAlz3} making it promising in the detection of preclinical disease \cite{ParAlz1}$^{,}$\cite{ParAlz2}$^{,}$\cite{IbaPar}. Our understanding of brain function relating to these tasks is incomplete due to technical limitations in analysis of brain recordings and neuroscientific limitations in understanding. It is clear that overcoming the former can guide the advancement of the latter. Thus, we set out to apply a novel signal processing approach to healthy brain functioning in order to uncover unimpaired brain processes which can help to guide future studies in the clinical setting. To this end, we contrast Shape and Bind conditions across healthy young subjects wearing an Electroencephalogram (EEG) to probe working memory integrative function at a high temporal resolution. In this study, we look at a set of VSTM cognitive tasks distinguished by two sets of binary conditions- Shape or Bind objects displayed in the left or right side of the screen (hemisfield), performed by healthy young volunteers, see Fig. \ref{Taskfig}. The Shape or Bind conditions allows the probing of the working memory integrative function whilst the hemisfield condition allows the exploration of contralateral sensitivities in brain activity to task performance.

The evidence gathered to date with the VSTM binding test comes from behavioural \cite{ParAlz1}$^{,}$\cite{ParAlz2}$^{,}$\cite{ParAlz3}$^{,}$\cite{Brock2008} and neuroimaging studies \cite{ParSCB}. Only one study has used EEG to investigate the early impact of AD on the mechanisms supporting this memory function \cite{PiettoSub}. None of these studies have addressed the issue of network organization and brain connectivity as the likely mechanisms sub-serving this memory function in intact brains. Should the brain network approach developed here prove informative with regard to unimpaired binding functions carried out in VSTM, this will create new opportunities to further investigate the impact of AD on memory. This is particularly relevant if we consider that AD causes severe brain disconnection from its very early stages \cite{Delb2007}$^{,}$\cite{Delb2003}. Several studies \cite{Brock2008}$^{,}$\cite{Rhodes2016}$^{,}$\cite{Lur2011} have demonstrated that there is no impairment in VSTM task performance due to healthy ageing so we do not expect the functionality of the brain to be different in these tasks for healthy old people and healthy young people. Thus, the information provided in this study should not pose age-translational problems to future research in the clinical setting.

EEG recordings provide a unique opportunity to deepen our understanding of human brain function across a healthy lifespan and in diseases of the nervous system \cite{LopEEG}. In the clinical context, the low cost and portability of the EEG offers a strong feasibility for screening purposes. Pertinently, it can aid in the early detection of brain dysfunction associated to diseases which have an impact on the worldwide population, such as dementia \cite{DauDem}$^{,}$\cite{SnaDem}$^{,}$\cite{StraDem}. In a broader context, the high temporal resolution of the EEG presents a great opportunity to study the rapid interdependent processes which underlie cognition \cite{LopEEG}. Thus, the EEG provides an unparalleled matching of practicality and data richness for neurological diagnostics.

In order to analyse the EEG recordings of these cognitive tasks, which generally require dynamic interactions between different brain areas, we focus on functional connectivity estimated by network modelling (functional networks)\cite{SporBook}$^{,}$\cite{BullRev}. Functional networks derive from network science which is a broadly applicable framework for understanding systems of interdependently acting agents. It is increasingly used in studies of brain function where bivariate connectivity methods applied to pairs of channels permits the construction of graphs which, in turn, leads to the identification of functional networks \cite{BullRev}$^{,}$\cite{BassRev}$^{,}$\cite{StamRev}. Such an approach is unveiling the functional architecture of the human brain, helping us to understand its vulnerability to brain diseases \cite{SporBook}$^{,}$\cite{BriNet}. In our setting, each node of the graph corresponds to an electrode placed on the scalp and the connection strength is the correlation of the time-series recorded at the electrodes. 

Because EEG channels pick up weak electromagnetic activity propagated through many layers of tissue and bone, the reproducibility of EEG activity over a heterogeneous population is difficult to ascertain. To address this, we choose to analyse modules of the network defined by several channels placed over localised spatial regions of the scalp. In this way the exact placements of EEG channels with respect to the underlying subject physiology becomes less relevant and so by focusing on these modules a better sensitivity to consistent effects over many subjects can be expected.

Exploiting the temporal resolution of EEG recordings in functional brain networks is an open problem ripe for exploration which, importantly, requires theoretical solutions and advancements \cite{FalRev}$^{,}$\cite{PapRev}$^{,}$\cite{CalfMRI}. Because the reliability of connectivity information is dependent on the number of time samples used, constructing networks over very short time epochs is implausible, which has heretofore restricted the ability to study rapid functional dynamics. To overcome this problem we introduce Modular Dirichlet Energy (MDE), a novel methodology based on graph signal processing \cite{ShuGSP}.

Graph signal processing is a new theoretical branch which combines graph theory and signal processing  \cite{ShuGSP}$^{,}$\cite{SanGSP}. In this setup, each node is intrinsically associated with some measurement or value regarded as a sample of a graph signal. This signal is not temporal but topological, so that the graph edges correspond to the topology over which the signal is supported \cite{ShuGSP}. The applicability of such methodology to the EEG is obvious since, during its recording, numerous channels are simultaneously sampled over the scalp. A weighted graph is generated using connectivity analysis between signals and we take the graph signal as the EEG recording at each electrode, hence the time-series, recorded at the electrode, is associated to the corresponding node. This allows two levels of analysis. The connectivity information encodes stable signal dependencies over long windows. This information then acts as a filter for temporal analysis provided by the graph signals over a short window, i.e. a retrospective temporal analysis of important connectivity. To study graph signals for specific modules, we set out the theory for MDE which combines Dirichlet energy, a measure of variability for functions, as it relates to graph signal processing \cite{ShuGSP} with the concept of modules from network science \cite{NewMod}.

Based on recent fMRI \cite{ParSCB} and EEG studies \cite{PiettoSub}$^{,}$\cite{ParraSub} we predict significant involvement of modules which map onto the functional network of working memory. Posterior regions (parietal-occipital) have been reported as important nodes of the network supporting VSTM binding tasks \cite{ParSCB}. Additionally, Parra et al. \cite{ParraSub} and Pietto et al. \cite{PiettoSub} recently reported involvement of frontal regions previously unnoticed by fMRI but known to be relevant to binding functions carried out in working memory \cite{Prab2000}. These authors argued that due to the high temporal resolution of the EEG, transient brain activity occurring in short time windows may be better detected by electrophysiological rather than fMRI techniques.  Regarding laterality Parra et al. \cite{ParraSub} recently reported hemispheric asymmetries during the retrieval phase in EEG while another study\cite{ParSCB} showed left hemisphere lateralisation in fMRI. Our methodological approach thus aims to unveil the network organisation that underpins such discrepancies by analysing transient dynamics of frontal and occipital modules. Furthermore we hypothesise that these regional activities are parts of an interdependent functional process and so we predict a specific difference in the activity between these modules which we analyse with Between MDE (BMDE).

\section*{Methods}
Our methodological approach contrasts network analysis of Shape and Bind conditions at two levels- modular weights of long-term task-related epochs and MDE of short-term windows. The first level uses a standard analysis of connectivity weights in order to discern which epochs (encoding and maintenance), conditions (Left and Right) and modules (Frontal and Occipital) reveal sensitive functional differences between Shape and Bind tasks. We then probe those discoveries for transient temporal dynamics by implementing the novel MDE analysis over 20ms windows. This ensures a rigorous process of discovery which can be easily replicated for other research questions. Fig.\ref{SPGpic}.A. outlines the main methodological steps of this study.

\subsection*{Visual Short-Term Memory Tasks and Data Acquisition}

\textit{Stimuli}: The stimuli were non-nameable shapes and non-primary colours known to be difficult to rehearse verbally \cite{ParAlz1}$^{,}$\cite{ParSCB}. 

\textit{Procedure}: Two arrays of three items each were presented to the left and to the right of a fixation cross centred on the screen on a grey background (Fig. \ref{Taskfig})\cite{Lur2011}. Each array was presented in a virtual 3x3 grid, 4$^{\circ}$ horizontally and 8$^{\circ}$ vertically centred and 3$^{\circ}$ to the left and right from fixation. Each item held 1$^{\circ}$ and the distance between items was never less than 2$^{\circ}$. Items for the study display were randomly selected from a set of eight polygons and eight colours \cite{ParAlz1} and randomly allocated to 3 of the 9 positions within the grid. At test the items randomly shuffled within the same locations used in the study display. Hence, items were never presented in the same locations across study and test displays. This rendered location uninformative.

Trials were self-initiated. A fixation cross appeared in the centre of the screen and remained on throughout the trial. After a button press, 500ms lapsed before the arrow clues were presented. Two arrows appeared for 200ms one above and one below fixation which indicated which of the two visual arrays (left or right) the participants were to attend. An interval of random duration selected from 400, 500 or 600ms followed the cues. The study display was then shown for 200ms. After an unfilled retention interval of 900 ms the test display appeared and remained visible until the participant responded.

In the Shape condition each array of the study display presented three black shapes. The test display also showed three shapes. In 50\% of the trials the content of the test display matched the content of the study display (“same trial”). The test display for the “different trials” showed two new shapes. In the Bind condition each visual array consisted of three shapes in different colours. In the test display for the “different trials” two coloured shapes swapped their colours. The participants responded “same” or “different” by pressing two keys previously allocated with both hands. The participants completed 8 practice trials before undergoing 170 test trials for each of the conditions.

Each participant undertook four different conditions of the VSTM task which are distinguished by two different binary manipulations- $2^{2}=4$. 
\begin{enumerate}
	\item Shape or Bind: the test items consist of black shapes (shapes only) or shapes with colours (shape-colour binding).
	\item Left or Right: the test items are shown on the left side or the right side of the screen (or hemisfield) to which the participant is prompted before stimulus onset.
\end{enumerate}
The task was to detect whether or not a change occurred across two sequential arrays shown on an initial study display and a subsequent test display.

EEG signals were recorded for 23 healthy young volunteers while they performed these VSTM tasks. Five of the volunteers were left-handed and eight were women. The mean and standard deviation (M $\pm$ SD) of the age of participants and number of years of education is 23.0 $\pm$ 4.3 and 17.1 $\pm$ 2.8, respectively. Informed consent was obtained from all subjects. The study was approved by the Psychology Research Ethics Committee, University of Edinburgh, and methods in data collection were carried out in accordance with their guidelines.

The EEG data was collected using NeuroScan version 4.3. The EEG was sampled at 250 Hz. A bandpass filter of 0.01-40 Hz was used. Thirty EEG channels, corrected for ocular artefacts using ICA, were recorded relying on the 10/20 international system. Fig. \ref{eegmap} shows the electrodes used in the analysis.

Further artefact rejection was conducted, rejecting trials which contained magnitudes of voltage fluctuations above 200 microvolts, transients above 100 microvolts and electro-oculogram activity above 70 microvolts. Only the trials with correct responses were kept as incorrect responses do not inform on working memory load in task comparisons. It is important to emphasise the distinction between a study of healthy brain function of task performance, as conducted here, for which the number of correct trials is not indicative, and a study of the performance of tasks by healthy people, for which the number of correct trials is indicative. In a few cases, no useful data was available for a volunteer performing one of the conditions resulting in an unequal number of volunteers per condition. 

To keep comparisons straightforward, we chose only to look at those 19 participants of the original 23 for whom data on all of the conditions was available. We focused on the encoding (i.e., study display) and maintenance (delay) periods of VSTM, since these seem to be the stages of memory informing about the functional principles of organisation with regard to capacity and format of representation (Shape vs Bind) \cite{ParSCB}. 

The mean $\pm$ standard deviation over participants for the number of kept trials for each condition were as follows: Shape, Left hemisfied- $69.74\pm 6.67$; Bind, Left hemisfield- $63.79\pm 8.72$; Shape, Right hemisfield- $66.32\pm 15.06$; Bind, Right hemisfield- $63.58\pm 16.26$.

\subsection*{Modular Dirichlet Energy}
Modular Dirichlet Energy, as we present it in this paper, is concerned with uncovering temporal dynamics of EEG channels associated with significant functional connectivity in an unprecedented and robust way. The technique considers the differences in amplitude recorded at EEG channels and weights them by the correlation between those channels leading to a metric that accounts for the contrast of amplitude and topological connectivity of brain activity. In a more intuitive sense, we can relate the graph connectivity as a filter to explore important areas for temporal dynamics in the signals. 

\begin{figure}[!t]
	\centering
	\includegraphics[trim = 0 200 0 200,clip,scale = 0.5]{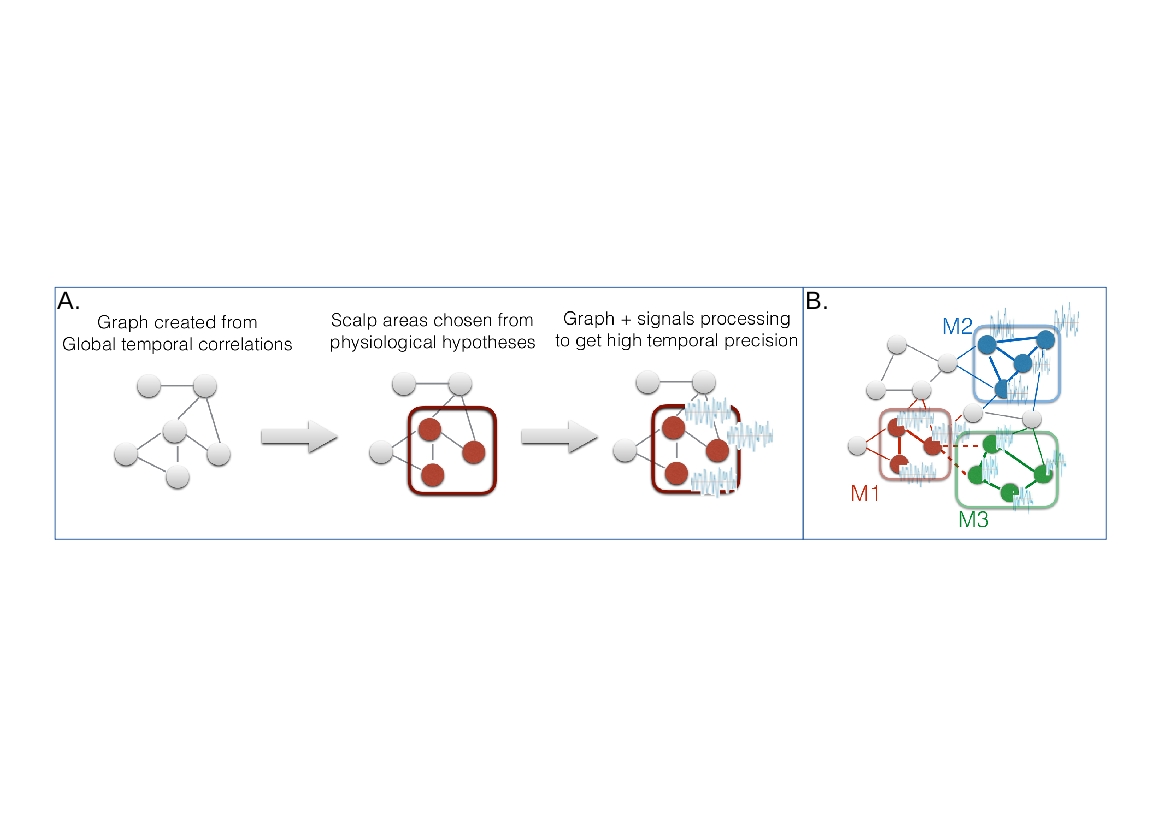}
	\caption{A. An outline of the main principles of the methodology. Circles represent electrodes and lines are the connections computed for the global time correlation. B. Example of modules for the Modular Dirichlet Energy (MDE). A set of electrodes are grouped together in modules (M1, M2, M3) within the network. The colored nodes and edges are the ones belonging to a specific module. The Between MDE of M1 (in red) and M3 (in green) is computed from the nodes connected by the dashed red edges. The Between MDE of M1 and M2 is zero since no edge connect these two modules.}
	\label{SPGpic}
\end{figure}

Let $G = (\mathcal{V}, \mathbf{f}, \mathcal{E}, \mathbf{W})$ be the mathematical representation of an undirected, labeled graph where $\mathcal{V}=\{1,\dots,n\}$ is the vertex set of the graph, $\mathbf{f}= \{f_{1},f_{2},\dots,f_{n}\}$ is the graph signal, or vertex amplitudes, indexed by  $\mathcal{V}$, $\mathcal{E} = \{ (i,j) \text{ s.t } i \text{ is adjacent to } j$ for $i,j\in \mathcal{V} \}$ is the edge set with $|\mathcal{E}|=2m$ and 
\[
\mathbf{W} = \left\{
\begin{array}{cc}
w_{ij} & (i,j)\in \mathcal{E}\\
0 & \text{ otherwise }\\
\end{array}
\right.
\]
is the weighted adjacency matrix of edge weights indexed by $\mathcal{E}$, where $w_{ij}$ is a measure of strength of relationship between nodes $i$ and $j$. Then we call $(\mathcal{V}, \mathbf{f})$ the vertex space of the graph and $(\mathcal{E}, \mathbf{W})$ the edge space of the graph so that $G$ is a dual space composed of the vertex and edge spaces where every element of the edge set in the edge space is an ordered pair of elements from the vertex set in the vertex space.

A module of a graph is defined by a subset of nodes and all adjacent edges to those nodes, i.e. $(\mathcal{V}_{X},\mathcal{E}_{X})$ where $\mathcal{V}_{X}\subset\mathcal{V}$ and $\mathcal{E}_{X} = \{ (i,j): i\in \mathcal{V}_{X}, j\in \mathcal{V}\}$. Likewise, the modular weights constitute the set $\{w_{ij}:  i\in \mathcal{V}_{X}, j\in \mathcal{V}\}$ and we define the total modular weight, $w_{\mathcal{V}_{X}}$, as
\begin{equation}\label{modweight}
w_{\mathcal{V}_{x}}=\sum_{i\in \mathcal{V}_{x}}\sum_{j\in\mathcal{V}}|w_{ij}|,
\end{equation}
for module $\mathcal{V}_{x}$, where the modulus indicates the strength of correlation relevant to our study. This allows us to probe the weighted connectivity information of the modules of relevant brain regions in the context of the whole network. For correlation, the higher the connectivity of the module, the more similarly the activity recorded at the modules signals are behaving with the entire graph and vice versa.

An example of a graph with modules and a graph signal is shown in Fig.\ref{SPGpic}.B. The circles represent the nodes of the network and the lines connecting them represent the edges. The graph signal is composed of the node amplitudes which are represented in Fig.\ref{SPGpic}.B by the EEG activity at the nodes. The full mathematical formulation of MDE and derivation of the formula for its computation is given in the supplementary material.

Suppose we have task-related activity in the brain with activity associated with module M1 and then transferring to module M3 over time. How do we analyse this activity in a way which can reveal the underlying chronological dependence of these two modules? After computing the connectivity of the signals over a suitable epoch, we are provided with two sets of information: the original high temporal resolution signals and the connectivity weights. The question we are then confronted with is how to use the high temporal resolution signals to explore temporal dynamics of the connectivity information. Graph Signal Processing provides the framework to answering this question.

We consider the relation
\begin{equation}\label{DE}
w_{ij}(f_{i}-f_{j})^{2}
\end{equation}

\noindent where $w_{ij}$ is the correlation between signals $i$ and $j$ over epoch $T$ and $f_{k}$ is the signal $k$ at time $t$ in $T$. When correlation between signals is low, the magnitude of \eqref{DE} is small and we cannot infer much from the signals. Now we consider when the magnitude of the correlation is large. There are four cases to consider. i) $w_{ij}>0$ and $(f_{i}-f_{j})$ is large; ii) $w_{ij}>0$ and $(f_{i}-f_{j})$ is small; iii) $w_{ij}<0$ and $(f_{i}-f_{j})$ is large; iv) $w_{ij}<0$ and $(f_{i}-f_{j})$ is small.

The first case says there is a strong positive correlation, $w_{ij}$, between signals $i$ and $j$ and $f_{i}$ and $f_{j}$ at time point $t$ are dissimilar. Since high positive correlation denotes that the signal amplitudes behave similarly, this large positive value indicates a likely discrepancy between the connectivity information over the epoch $T$ and the behaviour of the signals at time point $t$. Here the output is positive and large. The second case, on the other hand says that there is a strong positive correlation between the signals $i$ and $j$ and the signals are similar at $t$. This, in contrast to the first case, gives two agreeing components and the output is positive and small. The third case says there is a strong negative correlation, $w_{ij}$, between signals $i$ and $j$ and the magnitude of $(f_{i}-f_{j})$ is large. Since high negative correlation denotes that the signal amplitudes behave dis-similarly, this large positive value indicates agreement between the connectivity information over the epoch $T$ and the behaviour of the signals at time point $t$. The output is negative and large. The fourth case, on the other hand, says that there is a strong negative correlation between the signals $i$ and $j$ and the signals are similar at $t$. This, in contrast to the third case, gives two disagreeing components and the output is negative and small.

Consider summing a variety of these components, as in \eqref{dirichlet}. If the result is large and positive, we can say that the activity of the signals at time t is mostly in disagreement with the connectivity over $T$. On the other hand, if the output is large and negative, we can say that the activity of the signals at time $t$ is mostly in agreement with the connectivity over $T$. It follows that a value near 0 indicates activity which neither strongly agrees nor disagrees with the connectivity information.

Suppose now that $w_{ij}$ is significantly larger in condition $R$ (e.g. shape only) than in condition $S$ (e.g. shape- colour binding). Over a given population, we can generally expect that $w_{ij}(R)(f_{i}(R)-f_{j}(R))^{2} > w_{ij}(S)(f_{i}(S)-f_{j}(S))^{2}$. However, where this is most apparent will be where there is abnormal activity occurring in one of the conditions over the signal, thus indicative of a 'driving effect' for the correlation of a point in time in which the connectivity combined with the signal amplitudes is particularly important. We can extrapolate on this effect from the shape of the MDE throughout the duration of the connectivity window. If the driving effect occurs at a point where the MDE of one of the conditions undergoes a considerable change, i.e. the curve over time of the MDE is less smooth, we can infer that that condition shows a change in effect not noted in the other condition. If both conditions show contrasting changes, i.e. both curves are less smooth in contrasting directions, we can infer that the conditions take contrasting responses at that time. 

It is important to emphasise that if we analyse a sum of weights, $\sum_{i,j}w_{ij}$, the initial connectivity analysis should consider absolute values of correlation, where we are interested only in whether the correlation is strong (large magnitude) or weak (small magnitude), whereas the graph signal approach is most physiologically interpretable by considering the signs of the correlations, as above. 

Now, the Dirichlet energy of the graph $G$ is defined as 
\begin{equation}\label{dirichlet}
E(G) = \sum_{i,j=1}^{n}w_{ij}(f_{i}-f_{j})^{2}, 
\end{equation}
which is an inverse measure of the smoothness of the graph signal $\mathbf{f}$ over $G$ \cite{ShuGSP}. Particularly, the node gradient at node $i$ is an important measure of the smoothness of the graph signal at node $i$ and is defined as
\begin{equation}\label{nodegradient}
E(i) = \sum_{j}w_{ij}(f_{i}-f_{j})^{2}.
\end{equation}

If we consider this in our application, we see that the node gradient gives a local measure of the differences in the node's amplitude with respect to the rest of the electrode array 'filtered' by the strength of connectivity. For seeking well chosen modules, the node gradients can thus be used to ascertain which nodes' metrics are of comparable magnitude to be considered for the same module.

Note that the elements in the sum of \eqref{dirichlet} have a one to one mapping to the edge set, $\mathcal{E}$, of $G$. It follows that there is a natural decomposition of the Dirichlet energy in $\eqref{dirichlet}$, corresponding to any disjoint composition of the underlying graph into modules, such that
\begin{equation*}
E(G)= \sum_{x=1}^{M}\sum_{i\in \mathcal{V}_{x}}\sum_{j\in\mathcal{V}}w_{ij}(f_{i}-f_{j})^{2},
\end{equation*}
and we define the Modular Dirichlet Energy (MDE) of $\mathcal{G}_{x}$ to be
\begin{equation}\label{MDE}
MDE(\mathcal{G}_{x})  = \sum_{i\in \mathcal{V}_{x}}\sum_{j\in \mathcal{V}}w_{ij}(f_{i}-f_{j})^{2}.
\end{equation}
The red lines in Fig.\ref{SPGpic}.B represent all the edges, and corresponding Dirichlet energy components, of the module M1. Further, we can define the between module Dirichlet energy (BMDE) as
\begin{equation}\label{between}
BMDE(\mathcal{G}_{x},\mathcal{G}_{y})   = \sum_{i\in \mathcal{V}_{x}}\sum_{j\in \mathcal{V}_{y}}w_{ij}(f_{i}-f_{j})^{2}, 
\end{equation}
for two disjoint modules $\mathcal{G}_{x}$ and $\mathcal{G}_{y}$. In Fig.\ref{SPGpic}.B, the dashed red lines represent the between module edges and energy components of modules M1 and M3.

For graph signals which also have a temporal dimension, i.e. $\mathbf{F} = [\mathbf{f}^{0},\mathbf{f}^{1},\dots,\mathbf{f}^{Y}]$, an $n\times Y$ matrix of chronologically ordered graph signals $\mathbf{f}^{i}=\{f^{i}_{1},f^{i}_{2},\dots,f^{i}_{n}\}$, the Dirichlet energy of the signal during time period $[t_{0},t]$ is just the sum of the individual Dirichlet energies at each point in time. Thanks to linearity, this extends straightforwardly to all definitions above. By looking at short intervals of graph signals, $[t_{0},t]$, we can study moments in time of the network behaviour by looking at the graph signal in the short interval acting over the graph defined by connectivity of the whole epoch and thus probe the connectivity information for dynamic behaviour within the epoch on which the graphs are constructed.

MDE analyses temporal brain networks from a completely different angle to other state-of-the-art methods such as temporal networks \cite{Holme2012} or time series analysis of network metrics \cite{Sik2016}. Rather than being based on the construction of different networks indexed by chronology, MDE constructs just one network of general connectivity patterns over a larger epoch and uses this network as the support for localised time-series analysis of shorter windows. The activity is encoded in the graph signal rather than in the edge weights of a time-varying graph. This allows for contrasting the volatile behaviour of the EEG in a very short window (the graph signals) against the more stable activity computed over the long window (the weights of the graph edges), thus drawing out the temporal locations of driving effects of the connectivity differences whilst remaining robust to noise. This is arguably more elegant and directly comparable for transient dynamics than probing relations of several different networks corresponding to staggered time windows. Further, MDE works on weighted networks, as are ubiquitously generated by functional connectivity analysis, as well as binary networks, whereas multi-layer network approaches rely minimally on some threshold criteria of the edge weights.

Given that the components of the total modular weight are exactly those weights corresponding to the MDE of a given module, that the MDE is dependent on the underlying long-term modular edge weights and that the windows chosen for MDE are arbitrary, we recommend a two-level analysis approach. In the first level the data is probed over a long-term window using total modular weights. In the second level, analysis of short-term windows using MDE is implemented, see Fig.\ref{SPGpic}.A. We demonstrate this approach in this study.

\subsection*{Pre-processing and module selection}
The 30 channels were re-referenced to the average EEG activity. From the continuous EEG, we extracted epochs of 1.2 seconds starting at -200ms pre-stimulus onset (baseline). These epochs contained activity associated with the encoding and maintenance periods of the VSTM tasks. To further remove artefacts, channels with activity of a mean amplitude greater than 30 \textmu V (2 SD) were rejected. We then computed the average ERP signal over correct trials (number of correct trials per participants, per condition: mean- 65.7, SD- 9.27) for each VSTM condition performed by each participant resulting in a set of 4$\times$19 thirty-channel EEG signals.
\begin{figure}[t]
	\centering
	\includegraphics[trim = 40 430 0 80,clip, scale = 0.6]{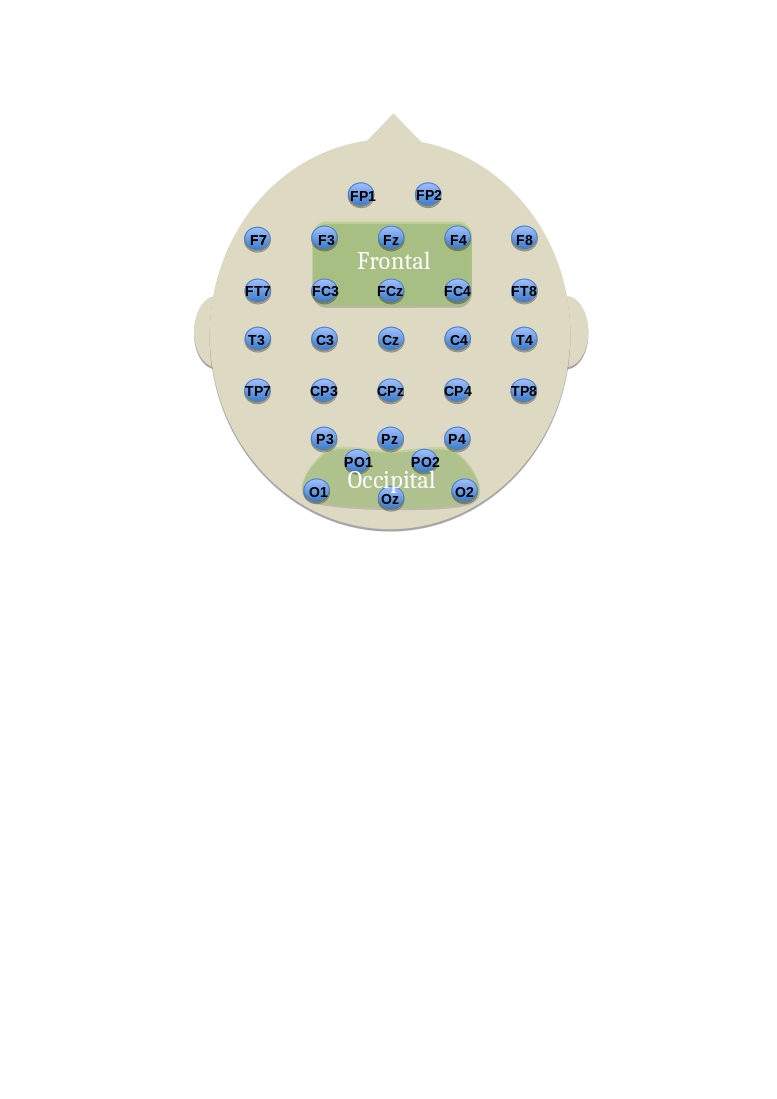}
	\caption{The frontal and occipital modules defined for network analysis. Labelled electrodes follow the 10-20 system.}
    \label{eegmap}
\end{figure}

We focused on activity in two time windows, one reflecting the encoding period (0-200ms) and one the maintenance period (200-1000ms). For each of the set of mean 30 channel signals, a graph was created for both time windows where the electrodes were mapped one to one with nodes and the edge weights were defined as the absolute value of correlation coefficient between the pairwise channels for the broadband of frequencies (0.01-40Hz). The correlation co-efficient is chosen, rather than phase-dependent connectivity, in order to analyse amplitude related effects from ERPs widely recognised as important in cognitive tasks. The broadband was considered to reflect the real-time amplitudes important to our novel analysis and in keeping with our processing-light approach. Further, a previous study of ERP broadband analysis on these tasks gave promising results \cite{PiettoSub}. For transparency, analysis of Theta, Alpha and Beta frequency bands can be found in the supplementary material. We also set the node amplitudes as the signals zero averaged over the channel space for each time sample.

In order to find differences in cognitive task conditions that are representative of the sampled population (e.g., controlling for sources of individual variability such as head size, small electrode displacements, etc.), we considered activity over broader regions involving several electrodes (mapping to modules in graph theory). To this aim, we defined two modules on the graphs using well known brain regions (see Fig. \ref{eegmap}) which are relevant to working memory processing \cite{ZimWMP} and previously reported to be involved in the task investigated \cite{ParSCB}$^{,}$\cite{Par2015}. These are the frontal module (F3, Fz, F4, FC3, FCz and FC4) and the occipital module (O1, Oz, O2, PO1 and PO2). To avoid combinatoric issues, these modules were chosen on physiological principles to be symmetric, of comparable size and with considerations of locality and generality in mind. This method is distinct from the algorithmic methods of finding modules based on the maximisation of the modularity metric\cite{NewMod}. Such a method constructs modules from the information of connectivity inside the signal with which one would instead seek to compare the composition of the found modules. Brain activity is dynamic and changes over time and space, thus such modules also change over time and space. The choice of modules we make is from physiological hypotheses based on previous knowledge. Defining a-priori modules on the graph is a way to combine these hypotheses with the information inside the signal, creating a topology of potential similarity between electrodes. The purpose of which is to provide a straightforward method to extract the dynamic information contained in the signal, as illustrated in Fig.\ref{SPGpic}.A.

The composition of the modules (i.e., electrodes chosen) was constructed after considering the node gradient \eqref{nodegradient} computed for each node of the graph during the entire encoding and maintenance period to determine their suitability for the modules (i.e., using energy to identify outliers). The occipital module was chosen considering all those electrodes in the occipital region. This choice is evidentially justified by the node gradients where there is clearly far larger magnitude gradients than in the rest of the electrode array (see Table \ref{nodeint}, column 5). In choosing the frontal module we wish to consider a comparable size of module to the occipital. If we consider the suitability of electrodes FP1 and FP2 for the frontal module, we see these electrodes have node gradients over 2 SD above the mean drawn from the rest of the electrode array excluding the occipital electrodes which exhibit obviously stronger
values (-18.56 $\pm$ 5.97 mean $\pm$ SD) thus we excluded them to avoid their overpowering influence since this highlights a strong contrast in activity. We then seek to form a symmetric module of comparable size to the occipital region, which leaves \{F3, Fz, F4, FC3, FCz and FC4\} as the physiologically feasible choice (Fig.\ref{eegmap}). Normalisation to correct for such influences is neither obvious nor advisable since each edge in a graph corresponds to two nodes and any such process would act to negate the scale-free nature of the underlying EEG network topologies \cite{SmiCWN}.

\begin{table}[t]
	\caption{Average node gradient \eqref{nodegradient} over task conditions and participants for each electrode}
	\label{nodeint}
	\begin{center}
		\begin{tabular}{|c|c||c|c||c|c||c|c||c|c|}
			\hline
			\textbf{FP1} 	&-32.45	&\textbf{F8}	&-19.33 	&\textbf{T3}	&-11.74 	&\textbf{CP3}	&-10.27	&\textbf{P4}   	&-30.00\\
			\hline
			\textbf{FP2}	&-32.07	&\textbf{FT7}	&-15.5	&\textbf{C3}	&-10.36	&\textbf{CPz}	&-13.60	&\textbf{O1}	&-59.71\\
			\hline
			\textbf{F7}	&-23.67	&\textbf{FC3}  &-17.66	&\textbf{Cz}	&-14.00	&\textbf{CP4}	&-16.01	&\textbf{Oz}	&-54.68\\
			\hline
			\textbf{F3} 	&-23.96	&\textbf{FCz} 	&-20.21	&\textbf{C4}	&-12.81	&\textbf{TP8} 	&-26.07	&\textbf{O2}	&-55.89\\
			\hline
			\textbf{Fz}  	&-28.09	&\textbf{FC4}	&-16.09	&\textbf{T4}	&-14.75	&\textbf{P3}	&-25.70	&\textbf{PO1}	&-56.29\\
			\hline
			\textbf{F4}   	&-20.32	&\textbf{FT8}  &-13.45	&\textbf{TP7}	&-17.45	&\textbf{Pz} 	&-26.32	&\textbf{PO2}	&-46.65\\
			\hline
		\end{tabular}
	\end{center}
\end{table}
We investigated differences in the encoding period (0-200ms) and the maintenance period (200-1000ms) of the tasks by analysing the total modular weight, $w_{\mathcal{V}_{x}}$\eqref{modweight}, of the specified modules.

\begin{figure}[t]
	\centering
	\includegraphics[trim = 40 275 0 75,clip,scale = 0.45]{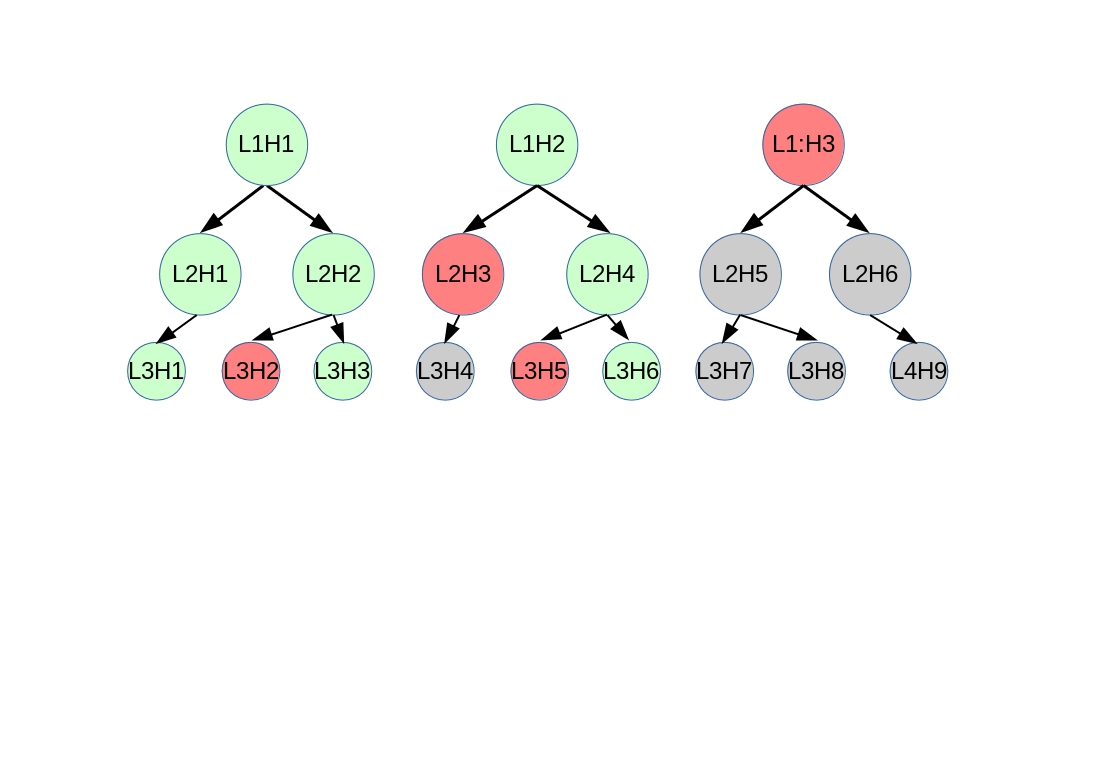}
	\caption{Example of hierarchical hypothesis tree for hierarchical false discovery rate procedure. 'L$i$H$j$' indicates the $j$th comparison in the $i$th level of the hierarchy. Red indicates no discovery, green indicates valid discovery, grey indicates exclusion from correction procedure due to false or no discovery made.}
    \label{hypotree}
\end{figure}

We contrast these values for Shape vs Bind conditions in the left hemisfield and in the right hemisfield using paired $t$-tests. Implementing the novel MDE (\ref{MDE}) concepts, where the graph signal is implemented over a reduced number of samples within the epochs, we introduce a second level of analysis to discover if particular parts of the original epochs are driving the discovered effects. Given the clear hierarchical structure of the hypotheses, it is then necessary to use hierarchical False Discovery Rate (FDR) \cite{YekFDR} to control for Type-I errors. Hierarchical FDR follows a level by level procedure of false discovery detection where a parent-child relationship is evident between these levels. Only those hypotheses whose parents were accepted as true discoveries are considered in the next level. In our study, the parent hypotheses relate to the total modular weights and the child hypotheses relate to the MDE analysis. Fig.\ref{hypotree} shows a model of a hypothesis hierarchy and the principles of rejection and acceptance of discovery through the FDR corrective procedure. We implemented a strict FDR with $q=0.05$ throughout the procedure. \cite{YekFDR}.

From the effects found in the edge weight testing, we compute the MDE for the frontal and occipital modules and the BMDE between the frontal and occipital modules. Note that BMDE $(\mathcal{G}_{x},\mathcal{G}_{y})\subset$MDE$(\mathcal{G}_{x})$, so that we probe the modules specifically for the interaction of the frontal and occipital modules.

\section*{Results} 
We contrast Shape and Bind values throughout, therefore metrics are usefully presented as Shape-Bind which implies the difference of the given metric values between the Shape and Bind condition. For this reason, we present boxplots indicating where the 0 line is for the Modular weights in the frontal and occipital modules (Fig.\ref{interp}, top left), and the MDE and BMDE of respective modules (Fig.\ref{interp}, bottom row). A summary of the results at two levels of analysis is presented in Tables \ref{weightTable} and \ref{energyTable}. Paired $t$-tests were performed over participants for the measurements obtained for Shape and Bind conditions. The paired $t$-test is a one-sample $t$-test with mean 0 on the values $(X-Y)$  for paired observations $X$ and $Y$ across subjects. With respect to our study, $X$ is a network metric of Shape task activity and $Y$ of Bind task activity. Therefore the null hypothesis is that Shape and Bind values come from the same normal distribution, i.e. mean$(X-Y) = 0$. The alternative hypothesis is that Shape and Bind values come from normal distributions with different means, i.e. mean$(X-Y) \neq 0$. The normality of the distributions was tested for each paired $t$-test using the one-sample Kolmogorov-Smirnov test. No significant deviations from the normal distribution were found at the 5\% level. Thus, the normality assumption underlying the paired t-tests is reasonable. The subsequent $p$-values were controlled using hierarchical FDR \cite{YekFDR}, allowing powerful probing of shorter time epochs of the discoveries found. We report the following:

\begin{figure}[!t]
	\centering
	\includegraphics[trim = 40 0 0 0, clip, scale=0.5]{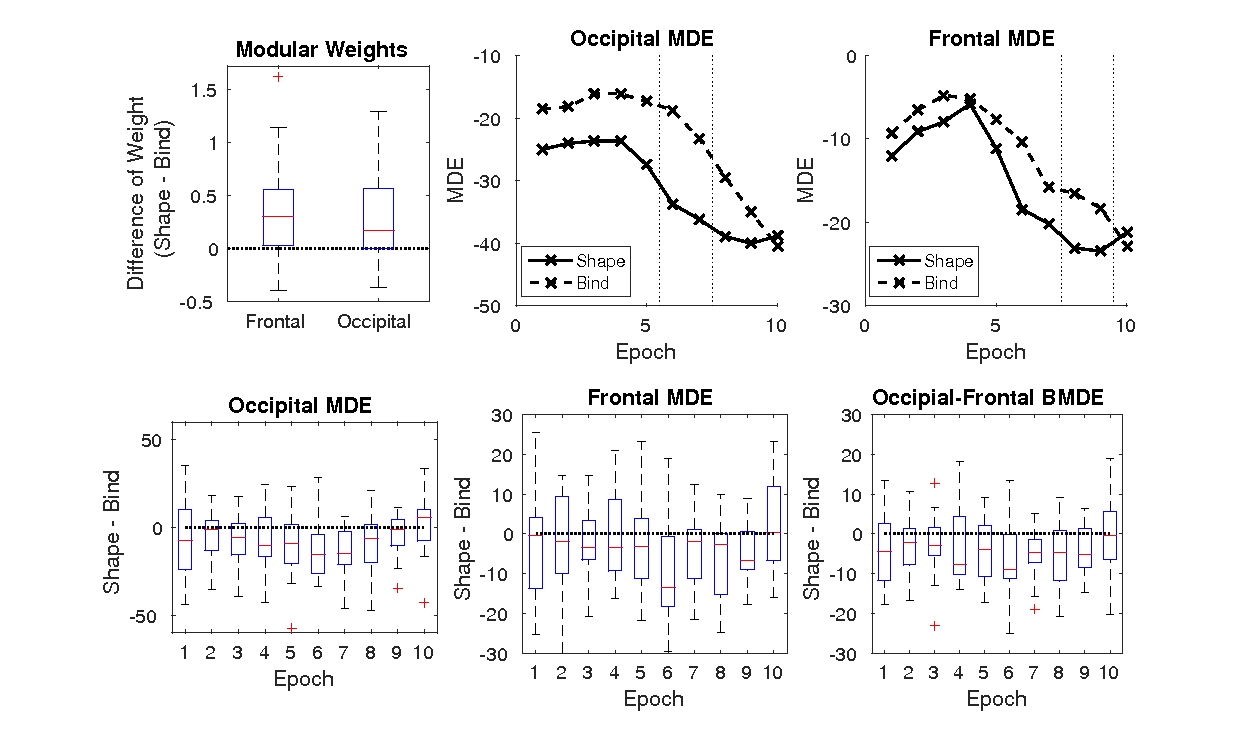}
	\caption{Top Left: difference in modular weights between the Shape and Bind tasks in the Frontal and Occipital modules. Top Center and right: Evolution (mean over subjects) of Modular Dirichlet Energy (MDE) of shape only (solid) and shape-colour binding (dashed) in the right hemisfield during the encoding period for the occipital module (center) and the frontal module (right) calculated over non-overlapping 20ms (5 time sample) windows. The dotted lines indicate the beginning and ending of the epochs displaying significant differences in activity. Bottom: Boxplots of MDE for each epoch for occipital (Left), frontal (Center) and the BMDE of Frontal and Occipital modules (Right), where 'Epoch' refers to the 20ms windows, labeled consecutively '1' to '10'.}
	\label{interp}
\end{figure}

\subsection*{Level 1}
In the first level, the long-term total modular edge weights computed from the absolute values of correlation are analysed for the conditions to be contrasted. These contrast are Left Shape vs Left Bind and Right Shape vs Right Bind in both frontal and occipital modules during both encoding and maintenance periods. Results found here thus inform on which periods, modules and task related hemisfields are important in Bind tasks. From the paired $t$-tests, after FDR correction, significant differences were found for the contrast involving Shape vs. Bind conditions in the right hemisfield (left hemisphere stimulation) for both frontal and occipital modules during the encoding period (see Table \ref{weightTable}). These showed that the Bind condition weights were less than those of the corresponding Shape conditions (Fig.\ref{interp}), top left. No differences were found in the maintenance period and, further, no differences were discovered when contrasting Shape vs Bind conditions in the Left hemisfield for either encoding or maintenance periods.

\subsection*{Level 2}
In the second level of analysis, we perform MDE over non-overlapping 20ms (5 time samples) windows over the modular weights (signed correlations) in Level 1. Due to the dependencies stated in the methods section, our analysis now focuses only on those hypotheses from which their parent hypothesis in the first level were seen as 'true discoveries'. Thus, we present results of the MDE for frontal and occipital modules during the encoding period of Shape vs Bind condition contrasts displayed in the right hemisfield. An extended table of results including those for the left hemisfield can be found in the supplementary material. Further, we study the BMDE of frontal and occipital modules to discover if there are epochs where dependencies occurring between these regions show strong effects.

After FDR correction, effects are found in the MDE of the occipital module in the epochs between 100-120ms and 120-140ms, showing a larger negative MDE in the Shape task. In the frontal module, an effect is found straight after this, between 140-160ms and 160-180ms, again showing a larger negative MDE in the Shape task. Further, the BMDE of occipital and frontal modules shows an effect in the epochs of 100-120ms and 120-140ms (see Table \ref{energyTable}). Notably, all the MDE values in this study were strong negative values indicating generally matching information between the signals and the connectivity weights, as explained in the methods. This is exactly as is expected, since the signals are those from which the connectivity information is taken.

\begin{table}[ht]
\centering
\caption{\label{weightTable}$p$-values for paired $t$-tests of modular sum of edge weights in Shape vs. Shape-colour binding conditions. O = occipital module, F = frontal module, E = encoding period, M = maintenance period, L = left hemisfield condition, R = right hemisfield condition. Blue = true discovery, red = false discovery, black = null hypothesis not rejected at the 5\% level.}
\begin{tabular}{|c|c|c|c|c|c|c|c|}
\hline
\textbf{O.E.L} & \textbf{O.M.L} & \textbf{O.E.R} & \textbf{O.M.R} & \textbf{F.E.L} & \textbf{F.M.L }& \textbf{F.E.R} & \textbf{F.M.R}\\
\hline
 0.1873 & 0.8709 & \color{blue}{0.0102} & 0.4514 & 0.2119 & 0.9040 & \color{blue}{0.0044} & 0.4806 \\
\hline
\end{tabular}
\centering
\caption{\label{energyTable}$p$-values for paired $t$-tests of Modular Dirichlet Energy (MDE) and Between MDE (BMDE) in Shape vs. Shape-colour binding conditions. Legend as in Table \ref{weightTable}.}
\begin{tabular}{|r|c|c|c|}
	\hline
	\textbf{Time (ms)} & \textbf{MDE - O.E.R }& \textbf{MDE - F.E.R} & \textbf{BMDE - F.O.E.R} \\
	\hline
	\textbf{0-20} & 0.2036 & 0.4088 & 0.0942\\
	\hline
	\textbf{20-40} & 0.0909 & 0.3891 & 0.0957\\
	\hline
	\textbf{40-60} & \color{red}{0.0432} & 0.1380 & 0.1408\\
	\hline
	\textbf{60-80} & 0.0718 & 0.8074 & 0.1805\\
	\hline
	\textbf{80-100} & \color{red}{0.0254} & 0.1918 & \color{red}{0.0412}\\
	\hline
	\textbf{100-120} & \color{blue}{0.0038} & \color{red}{0.0465} & \color{blue}{0.0073}\\
	\hline
	\textbf{120-140} & \color{blue}{0.0010} & 0.0851 & \color{blue}{0.0028}\\
	\hline
	\textbf{140-160} & \color{red}{0.0278} & \color{blue}{0.0070} & \color{red}{0.0120}\\
	\hline
	\textbf{160-180} & 0.0919 & \color{blue}{0.0059} & \color{red}{0.0167}\\
	\hline
	\textbf{180-200} & 0.6661 & 0.5464 & 0.9644\\
	\hline	
\end{tabular}
\end{table}

\begin{figure}[!t]
	\centering
	\includegraphics[trim = 0 0 0 0, clip, scale=0.35]{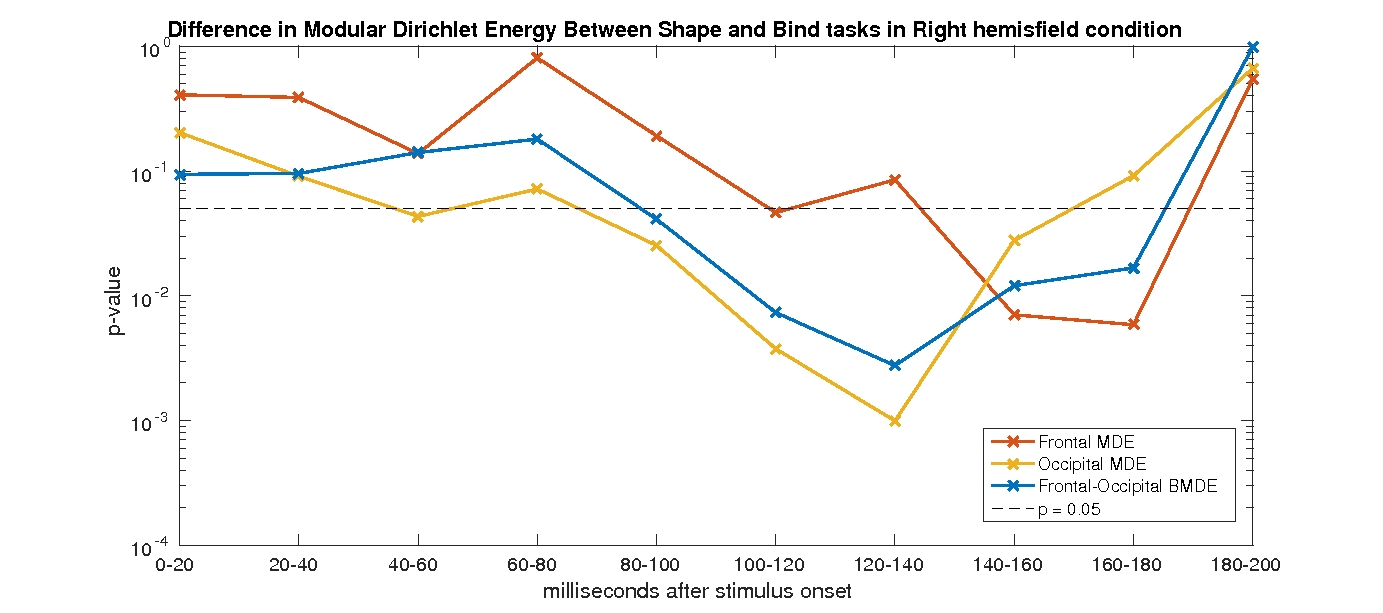}
	\caption{The $p$-values for shape only vs. shape-colour binding contrasts in the right hemisfield during the encoding period for the Modular Dirichlet Energy (MDE) in the occipital module (blue) and the frontal module (yellow) along side the Betweenness Modular Dirichlet Energy (BMDE) of the occipital and frontal modules calculated over non-overlapping 20ms (5 time sample) windows. The y-axis is on a logarithmic scale.}
	\label{semilogy}
\end{figure}

\section*{Discussion}
Agreeing with our hypotheses, we found evidence that the contrasting brain function of Shape vs Bind conditions occurs during the encoding period in both the frontal and occipital modules, which are regions classically associated with task performance and visual attention. Particularly, this supports the evidence found in Parra et al. \cite{ParraSub} and Pietto et al. \cite{PiettoSub} that these tasks involve rapid functional activity in the frontal module which is picked up by the EEG. The modular weights being generally stronger in the Shape task than the Bind task suggest that the activity in these regions is more in agreement with the rest of the network in the Shape task. The modular correlates being weaker in the Bind task indicates a greater contrast in activity between these regions and the rest of the network and thus is suggestive of this task placing a greater work load on these regions, in keeping with our expectations.

Importantly, we found that this activity was clearly evident in the right hemisfield condition but not so in the left hemisfield condition, providing evidence to suggest that there is better sensitivity of EEG activity in tests designed for the right hemisfield. Right hemisfield implies stimulation of the left brain hemisphere which is classically the more task-oriented half of the brain. From this we conjecture that a more direct route, and so less convoluted connectivity, is provided by left brain stimulation for task related activity. It follows that, for the most clinically relevant biomarker for AD by using these tasks in conjunction with the EEG, performance of tasks presented in the right hemisfield may provide more rich information and thus greater sensitivity to impaired activity than those presented elsewhere on the screen.

From the MDE, we note that the effects reported in the occipital module appear to be driven by amplitude based activity between 100-140ms into the encoding phase of the task. This coincides with the P100 of visual evoked potentials and shows that with our methodology we are able to pick up on event related potential (ERP) activity over the network, which was previously impossible in functional brain network studies because of connectivity values being necessarily computed from information over longer epochs. During P100, then, the Shape condition exhibited a noticeable dip in MDE which was much less apparent in the Bind case (Fig.\ref{interp}, top center). It is reasonable to suggest that this is caused by the greater work load in the Bind condition. The involvements of visual association cortices in regions of the occipital lobe during short-term memory binding has been documented previously \cite{ParSCB}. This appears to be a key area of the visual integrative functions. Recent studies using EEG based methods have confirmed the involvement of these regions in the poor performance found in patients at risk of AD \cite{PiettoSub}$^{,}$\cite{ParraSub}.

Recent electro-physiological studies indicate that frontal nodes may be contributing both specific (i.e. binding) and more general resources during working memory processing. The effect seen here between the frontal and occipital modules from 100-140ms and that seen in the frontal module between 140-180ms concurs with this, suggesting that a contrast exists in the functional dependency between these regions for Shape vs Bind conditions shortly after the onset of P100 activity. Further the activity occurring in the Frontal module indicates a difference in higher function post-visual processing. Here, the Bind and Shape conditions show contrasting MDE activity in these epochs where the dip in the Shape condition is contrasted by the flattening in the Bind condition (Fig.\ref{interp}, top right). It is reasonable to expect that this activity again relates to increased load of shape-colour binding, pointing reasonably to an inverse dependency of information load with functional efficiency. This information may inform us on the deficits found in the clinical setting. Alzheimer's disease is characterised by both deficits in new learning (which relates to the encoding period) and rapid forgetting, so that pinpointing spatio-temporal abnormalities in the biological substrates that underpin these deficits is essential both to understand key disease processes and as a step towards defining a useful biomarker of preclinical AD. On this basis, we conjecture that this occipital-frontal dependency is weakened in AD patients. It is reasonable to predict that this functional processing effect will be noticeable in future clinical studies and thus could eventually prove useful in providing an indicator for a sensitive biomarker.

The results suggest a focused prolonged functional difference between Shape and Bind conditions beginning in the occipital area at around 100ms, with a dependency between occipital and frontal areas from 100ms to 140ms and then shifting towards the frontal area between 140-180ms, see Fig. \ref{semilogy}. The strong chronological dependency of $p$-values over non-overlapping epochs is remarkable. Additionally, it is noticeable that all these effects have entirely vanished by the 180-200ms epoch, which is in accordance with the lack of findings found for the maintenance period.

In application, the MDE proves to be a sensitive and highly flexible methodology for EEG analysis, both topologically and temporally, providing a unique platform to study EEG activity, such as ERPs, as they apply to functional brain networks. In fact, we showed that not only can MDE pick up on well documented EEG activity, but it can extend our understanding of that activity as a dynamic interdependent activity between different brain areas, progressing our understanding beyond singular channel effects.

Technically, implementation of MDE and its related components could narrow to the level of single time samples, although in practice this would be difficult to justify for EEG recordings. However, as EEG technology improves and knowledge of brain function deepens, as shown here, this method has the potential for pinpointing shifts in functional brain dynamics to the millisecond. As with any such generalisable technique, the flexibility of this methodology means that the potential number of tests which can be carried out is huge. It is important then, as demonstrated here, to make a rigorous set of hypotheses before implementation. 

\section*{Acknowledgements}
This study was partially supported by the Engineering and  Physical Sciences Research Council (UK) via a DTP studentship to KS and the research project EP/N014421/1 to JE. KS was also awarded a JM Lessells Travel Scholarship from the Royal Society of Edinburgh to undertake collaborative research at EPFL. NS was supported by SNF grant 200021 154350/1 for the project ``Towards signal processing on graphs''. AI is supported by CONICET, CONICYT/FONDECYT Regular (1130920), FONCyT-PICT 2012-0412, FONCyT-PICT 2012-1309, FONDAP 15150012, and INECO Foundation. MAP work was supported by Alzheimer's Society, Grant \# AS-R42303. This study was also supported by the MRC grant \# MRC-R42552, awarded to MAP in collaboration with AI and JMS. We thank Jamie Crowther who assisted with data collection. We also acknowledge the support from the Alzheimer's Scotland Dementia Research and the Centre for Cognitive Ageing and Cognitive Epidemiology part of the cross council Lifelong Health and Wellbeing Initiative (MR/K026992/1) both from the University of Edinburgh. 

\section*{Author contributions statement}
MAP, AI, KS and JE formulated the hypotheses. KS, BR and NS developed the theory of MDE, supervised by PV. KS BR and JE conceived the methodology. MAP and JMS contributed to the discussion.  MAP and SR collected the data. KS performed the analysis of the data and wrote the manuscript. All authors reviewed the manuscript.

\section*{Additional information}
The authors declare no competing financial interests.

\end{document}